\documentclass[11pt,a4paper]{article}
\pdfoutput=1
\usepackage{jheppub}
\usepackage{amsfonts}
\usepackage{bbm}
\usepackage{graphicx}
\usepackage{caption}
\usepackage{subcaption}
\usepackage{bigints}
\usepackage[normalem]{ulem}
\usepackage{slashed}

 \def\g{\gamma}

\def\e{\epsilon}

\def\i{\iota}

\def\k{\kappa}    
    
\def\m{\mu}
\def\n{\nu}

\newcommand{\cD}{\mathcal{D}}

\newcommand{\cM}{\mathcal{M}}

\newcommand{\RN}[1]{%
  \textup{\uppercase\expandafter{\romannumeral#1}}%
}

\newcommand{\ndt}{\noindent}

\newcommand{\nn}{\nonumber}

\def\e{\epsilon}

\def\i{\mathrm{i}}

\def\p{\partial}

\def\bea{\begin{eqnarray}}
\def\eea{\end{eqnarray}}
\def\be{\begin{equation}}
\def\ee{\end{equation}}
\def\ba{\begin{align}}
\def\ea{\end{align}}

\newcommand{\bem}{\begin{pmatrix}}
\newcommand{\eem}{\end{pmatrix}}

\def\={\;  = \;}
\def\+{\, + \,}

\def\bar{\overline}

\def\rt2{\sqrt{2}}

\title{Supersymmetric Graphene on Squashed Hemisphere}

\author[a]{\small{Rajesh Kumar Gupta},}
\author[b]{\small{Augniva Ray}}
\author[a,c]{\small{and Karunava Sil}}
\affiliation[a]{\small{Department of Physics, Indian Institute of Technology Ropar,
Rupnagar, Punjab 140001, India}}
\affiliation[b]{\small{Theory Division, Saha Institute of Nuclear Physics, HBNI, 1/AF Bidhan Nagar, Kolkata, India - 700064}}
\affiliation[c]{\small{School of Basic Sciences, Indian Institute of Technology Bhubaneswar, Bhubaneswar 752050, India}}

\emailAdd{rajesh.gupta@iitrpr.ac.in}
\emailAdd{augniva.ray@saha.ac.in}
\emailAdd{ks45@iitbbs.ac.in}

\abstract{We compute the partition function of $\mathcal N=2$ supersymmetric mixed dimensional QED on a squashed hemisphere using localization. Mixed dimensional QED is an abelian gauge theory coupled to charged matter fields at the boundary. The partition function is a function of the complex gauge coupling $\tau$, the choice of R-symmetry and the squashing deformation. The superconformal R-symmetry is determined using the 3-dimensional F-maximization. The free energy as a function of squashing deformation allows computing correlation functions that contain the insertion of the energy-momentum tensor. We compute the 2-point correlation function of the boundary energy-momentum tensor by differentiating the free energy with respect to the squashing parameter. We comment on the behaviour of the 2-point function as we change the complex coupling $\tau$.
}

\makeatletter
\gdef\@fpheader{}
\makeatother
\begin{document}

%
%\vspace*{-2cm} 
%\begin{flushright}
%{\tt  KCL-MTH-16-07 } 
%\end{flushright}

\maketitle

\section{Introduction \label{sec:intro}}
Quantum field theory on a manifold with boundary finds many applications ranging from string theory to condensed matter physics such as D-branes, topological insulators and graphene. The fixed point in the renormalization group flow in quantum field theory is particularly interesting since a conformal field theory describes it. These theories find a useful application in condensed matter system, for example, in describing the second-order phase transition. 
Our focus here will be on 4-dimensional conformal field theories in the presence of a boundary with conformally invariant boundary conditions. One such example of 4-dimensional boundary conformal field theories (bCFT) is mixed dimensional quantum electrodynamics where 4-dimensional electromagnetic field interacts with charged matter fields on the 3-dimensional boundary. These theories exhibit many interesting properties, see for example~\cite{Hsiao:2017lch, Hsiao:2018fsc,DiPietro:2019hqe, Grignani:2019zxc}.

One of the characteristic features of a conformal field theory is the presence of a quantitative measure of the number of degrees of freedom that decreases along the RG flow connecting two CFTs. In 2 and 4 dimensions, it coincides with the central charge $c$ and $a$, respectively, while in 3-dimensional CFTs, the free energy of the theory computed on S$^3$ plays the similar role. Monotonicity theorem also exists in $d$-dimensional bCFTs~\cite{PhysRevLett.67.161,Friedan:2003yc, Jensen:2015swa, Casini:2016fgb,Casini:2018nym}. The boundary free energy defined from the hemisphere partition function as
\be
\frac{|Z_{HS^{d}}|^{2}}{Z_{S^{d}}}=e^{\text{div.}-2F_{\p}}\,,
\ee
where terms in ``div.'' are divergent terms which have $(d-1)$-dimensional origin, conjectured to decrease along the RG flow triggered by the boundary relevant operator. Our goal would be to compute the boundary free energy in the 4-dimensional supersymmetric bCFTs.

The localization computation of the partition function for 4-dimensional $\mathcal N=2$ supersymmetric theories with Neumann and Dirichlet boundary conditions was first performed in~\cite{Gava:2016oep}. 
Subsequently, the partition function of $\mathcal N=2$ supersymmetric graphene-like theories with general boundary conditions appeared in~\cite{Gupta:2019qlg}. In the paper~\cite{Gupta:2019qlg}, the authors computed the partition function as a function of trial R-charge of the charged matter fields at the boundary. The partition function takes the form of real one-dimensional integral with integrand given in terms of Jafferis $\ell$-function. One of the novel features of the partition function is that it depends on the complexified gauge coupling
\be
\tau=\frac{\theta}{2\pi}+\frac{2\pi i}{g^{2}}\,,
\ee
which is exactly marginal. This gives rise to a boundary conformal field theory for every value of the complex coupling $\tau$. 
Thus, the partition function computed using the method of localization is a function of the complex coupling $\tau$ and the choice of R-symmetry. The R-symmetry is determined using the 3-dimensional F-maximization~\cite{Jafferis:2010un}.
Moreover, the partition function, as a function of the background sources for gauge and flavor currents, was used to compute boundary transport coefficients that appear in the 2-point function of the corresponding currents.

In the present article, we will generalize the above computation to include the metric background deformations.  More specifically, we will consider $\mathcal N=2$ supersymmetric mixed dimensional QED on a squashed hemisphere. Supersymmetric theories on a squashed sphere have been studied in various dimensions. Provided the squashing deformations preserve some supersymmetry, the partition function as the function of the squashing parameters can be computed using the localization technique, see for example~\cite{Hama:2011ea,Imamura:2011wg, Hama:2012bg,Imamura:2012xg}.
The free energy as a function of squashing deformations allows us to compute correlation functions that contain the insertion of the energy-momentum tensor.
We will be interested here to compute the transport coefficient that appears in the 2-point function of the energy-momentum tensor in 3-dimensional boundary.

Conformal symmetry and the conservation law fix the 2-point function of energy-momentum tensor in 3-dimensional flat space up to a constant. It is given by
\bea
<T_{\mu\nu}(x)T_{\rho\sigma}(0)>=-\frac{\tau_{R}}{64\pi^2}(\delta_{\mu\nu}\p^2-\p_\mu\p_\nu)(\delta_{\rho\sigma}\p^2-\p_\rho\p_\sigma)\frac{1}{x^2}\nn\\+\frac{\tau_{R}}{64\pi^2}\Big((\delta_{\mu\rho}\p^2-\p_\m\p_\rho)(\delta_{\n\sigma}\p^2-\p_\n\p_\sigma)+(\m\leftrightarrow\n)\Big)\frac{1}{x^2}\,.
\eea
The quantity of interest $\tau_{R}$ can be computed by placing the conformal field theory on S$_{b}^{3}$. It is given by the second derivative of the squashed free energy with respect to the squashing parameter $b$~\cite{Closset:2012ru}
\be
\tau_{R}=\frac{2}{\pi^{2}}\text{Re}\frac{\p^2F_{b}}{\p b^2}\Big|_{b=1}, \qquad \text{where}\quad F_{b}=-\ln Z_{b}\,,
\ee
where the free energy is evaluated using the superconformal R-charge. 

We extend the above computation to 2-point function in bCFT. In particular, we compute the 2-point function of the boundary energy-momentum tensor by differentiating the boundary-free energy as
\be
\tau_{R}=\frac{2}{\pi^{2}}\frac{\p^2F_{\p,b}}{\p b^2}\Big|_{b=1}\,.
\ee
Basically, the idea is that after integrating over the bulk degrees of freedom for a given conformal boundary condition (and dividing by the sphere partition function), we can think of the $F_\p$ as the free energy of some effective CFT at the boundary. At the perturbative level, the effect of bulk degrees of freedom can be mimicked by introducing interactions in the boundary theory involving auxiliary fields. These were the original arguments in~\cite{Gaiotto:2014gha} that provided evidence in support of the decrease of boundary-free energy along the boundary RG flow.

To compute $\tau_{R}$, we follow the following strategy~\cite{Nishioka:2013gza}: 1) Evaluate the partition function on S$^{3}$ and determine the R-charge that maximizes the real part of the free energy, 2) Use the R-charge thus obtained to evaluate the free energy on S$_{b}^{3}$, and 3) Compute the second derivative of the free energy, which is the function of squashing parameter $b$, to obtain the expression for $\tau_{R}$.

The outline of the rest of the paper is as follows. In section 2, we review two different ways of squashing a 4-dimensional sphere. We then discuss supersymmetry on the squashed sphere. 
After that, we discuss the condition the background fields need to satisfy to have supersymmetry on the squashed hemisphere. 
In section 3, we find the partition function of mixed dimensional QED coupled to charged matter fields at the boundary.  In Section 4, we compute the 2-point function of the energy-momentum tensor. We then conclude with a summary in section 5. In Appendix A,
we present our conventions for gamma matrices and reality condition on fermions. In appendix B, we discuss the supersymmetry on the squashed hemisphere. Here we also discuss the requirement on the supergravity background fields. In appendix C and D, we present the supersymmetric action and background fields on the squashed sphere. In appendix E, we give explicit expressions for functions that appear in the subleading computations of $\tau_R$. 
\section{Supersymmetry on squashed hemisphere}
In this section, we will introduce the supersymmetric background on a squashed hemisphere. The squashed hemisphere is obtained by considering the squashed sphere and placing the boundary at the equator.
We will be considering here two different kinds of squashing depending on the isometry preserved by the deformation of S$^{4}$. 
The first kind of squashing preserves $SU(2)\times U(1)$ isometry. The metric is given by
\be\label{SquashedMetric1}
ds_{\RN{1}}^{2}=dr^{2}+\frac{1}{4}\sin^{2}r\Big(d\theta^{2}+\sin^{2}\theta\,d\phi^{2}+h(r)^{2}(d\psi+\cos\theta\,d\phi)^{2}\Big)
\ee
where $0\leq r\leq \pi$, $0\leq \theta\leq \pi$, $0\leq\phi\leq2\pi$ and $0\leq\psi\leq4\pi$ and $h(r)$ is an arbitrary smooth function. The regularity of the metric near the north pole (i.e. at $r=0$) and south pole (i.e. at $r=\pi$) requires the following behaviour the function  
\bea\label{SmoothCond}
&&\text{North pole:}\qquad h(r)=1+c_{2}r^{2}+\mathcal O(r^{3})\,,\nn\\
&&\text{South pole:}\qquad h(r)=1+d_{2}(\pi-r)^{2}+\mathcal O((\pi-r)^{3})\,.
\eea
The squashed hemisphere is obtained by the requirement that $0\leq r\leq \frac{\pi}{2}$. The induced metric at the boundary is given by
\be\label{SquashedMetric1}
ds_{\p\RN{1}}^{2}=\frac{1}{4}\Big(d\theta^{2}+\sin^{2}\theta\,d\phi^{2}+h\Big(\frac{\pi}{2}\Big)^{2}(d\psi+\cos\theta\,d\phi)^{2}\Big)
\ee
The above metric on the squashed S$^{3}$ preserves $SU(2)\times U(1)$ symmetry.
Thus, any arbitrary smooth function satisfying the regularity condition \eqref{SmoothCond} gives rise to a smooth deformation of the HS$^{4}$. The deformation can then be used to compute the 2-point function of energy momentum tensor of the boundary conformal field theory. For the convenience of the later computations, will choose the following form of the function 
\be\label{h}
h(r)=1+\frac{2\alpha}{\pi^{2}}\sin^{3}r\,.
\ee
In the above $\alpha$ is a positive real parameter with $\alpha=0$ correspond to unsquashed background. The above form of $h(r)$ satisfies the smoothness criteria given in~\eqref{SmoothCond}.\\

The second kind of the deformation we will consider preserves $U(1)\times U(1)$ isometry. The squashed metric is given by
\be\label{SquashedMetric2}
ds_{\RN{2}}^{2}=\delta_{ab}E^{a}E^{b}\,,
\ee
where the vielbeins are
\bea
&&E^{1}=\ell\, \sin r\,\cos\theta\,d\phi,\quad E^{2}=\tilde\ell\,\sin r\,\sin\theta\,d\chi,\quad E^{3}=f(\theta)\,\sin r\,d\theta+h(r,\theta)dr,\nn\\
&& E^{4}=g(r,\theta)\,dr\,.
\eea
In the above, the range of the coordinates are $0\leq r\leq \pi$, $0\leq \theta\leq \frac{\pi}{2}$ and $0\leq\phi,\chi\leq 2\pi$. The parameters $\ell$ and $\tilde\ell$ are constants and the functions appearing in the metric are given by
\bea
&&f(\theta)=\sqrt{\ell^{2}\sin^{2}\theta+\tilde\ell^{2}\cos^{2}\theta},\quad g(r,\theta)=\sqrt{\rho^{2}\sin^{2}r+\frac{\ell^{2}\tilde\ell^{2}}{f(\theta)^{2}}\cos^{2}r}\,,\nn\\
&&h(r,\theta)=\frac{\tilde\ell^{2}-\ell^{2}}{f(\theta)}\cos r\,\sin\theta\,\cos\theta\,.
\eea
The induced metric at the boundary of the squashed hemisphere i.e. at $r=\frac{\pi}{2}$ is 
\be
ds_{\p\RN{2}}^{2}=\ell^{2}\cos^{2}\theta\,d\phi^{2}+\tilde\ell^{2}\sin^{2}\theta\,d\chi^{2}+(\tilde\ell^{2}\cos^{2}\theta+\ell^{2}\sin^{2}\theta)\,d\theta^{2}\,.
\ee
Next, we will discuss the background which needs to be turned on in order to preserve supersymmetry on the squashed hemisphere.
\subsection{Supersymmetric background on squashed hemisphere}
We will be interested in $\mathcal N=2$ supersymmetric theory on the squashed hemisphere backgrounds \eqref{SquashedMetric1} and \eqref{SquashedMetric2}. The squashed metric background does not admit any rigid supersymmetry itself. However, the supersymmetric theory can be put on the squashed hemisphere if we turn on some non-dynamical supergravity background fields. The theory is invariant under the rigid supersymmetry transformations that are generated by the solution of the Killing spinor equation
\be\label{KillingSpEq1}
\mathcal D_{\m}\xi^{i}+T^{ab}\gamma_{ab}\gamma_{\m}\xi^{i}=\gamma_{\m}\xi'^{i}\,.
\ee
In addition to above, the Killing spinor also satisfies an auxiliary equation
\be\label{KillingSpEq2}
\slashed\cD\slashed\cD\xi^{i}+4\cD_{c} T_{ab}\g^{ab}\gamma^{c}\xi^{i}=M\xi^{i}\,.
\ee
The fields $T^{ab}$ and $M$ are auxiliary background fields. Also, the $SU(2)_{R}$ connection appears in the Killing spinor equation through the covariant derivative given by
\be
\mathcal D_{\m}\xi^{i}=(\p_{\m}+\frac{1}{4}\omega_{\m ab}\gamma^{ab})\xi^{i}+\mathcal V^{i}_{\m\,j }\xi^{j}\,.
\ee
These Killing spinors also satisfy the symplectic Majorana condition
\be\label{SymMajRealityCondition}
\bar\xi_{i}\equiv(\xi^{i})^{\dagger}=\epsilon_{ij}\xi^{j\,T}C\,,
\ee
where $C$ is the charge conjugation matrix.\\
Now, the presence of the boundary at $r=\frac{\pi}{2}$ breaks half of the supersymmetry transformations i.e the boundary in our case preserves 4 out of 8 supercharges. Given a solution to the Killing spinor equations \eqref{KillingSpEq1} and \eqref{KillingSpEq2}, $\xi^{i}$, we define the following projected spinors~\cite{Gupta:2019qlg}
\be
\xi^{i}_{\pm}=\Pi^{i}_{\pm\,j}\xi^{j}\,,
\ee
where the projector is given by
\be
\Pi^{i}_{\pm\,j}=\frac{1}{2}(\delta^{i}_{j}\pm i\tau^{i}_{3\,j}\gamma_{5}\gamma_{n})\,.
\ee
Here, $\gamma_{n}$ is the flat space gamma matrix corresponding to the direction perpendicular to the boundary i.e. $n$. We require that at $r=\frac{\pi}{2}$, $\Pi^{i}_{+\,j}\xi^{j}\Big|_{r=\frac{\pi}{2}}=0$, and $\xi^{i}_{-}\Big|_{r=\frac{\pi}{2}}$ generates the supersymmetry on the boundary. It is important to note that $\xi^{i}_{\pm}$ are not the solution of the 4-dimensional Killing spinor equations. \\
The spinor $\xi^{i}_{-}\Big|_{r=\frac{\pi}{2}}$ being the Killing spinor on the boundary, i.e. a solution to the boundary Killing spinor equation, requires the background to satisfy certain conditions. Using the fact that the components of spin connections corresponding to the metrics \eqref{SquashedMetric1} and \eqref{SquashedMetric2} satisfy
\be
\omega^{Bn}_{A}=0,\qquad\text{for}\quad A,B=1,2,3\,, 
\ee
it turns out that $\xi^{i}_{-}\Big|_{r=\frac{\pi}{2}}$ solves 3-dimensional Killing spinor equation if the background satisfies the following condition (see the Appendix \eqref{ProjectorAndKilling} for more details)
\be\label{antisyscon}
T_{Bn}|_{r=\frac{\pi}{2}}=0\,.
\ee
Note that the above condition is the requirement for having $\mathcal N=2$ supersymmetry with $U(1)_{R}$ symmetry at the boundary.
Thus, the spinor $\xi_{-}^{i}$ at the boundary satisfies the Killing spinor equation
\be
\cD^{3d}_{A}\xi_{-}^{i}+T_{BC}(\gamma_{BC}\gamma_{A}+\frac{1}{3}\gamma_{A}\gamma_{BC})\xi_{-}^{i}=\frac{1}{3}\gamma_{A}\gamma^{B}\cD^{3d}_{B}\xi_{-}^{i}\,.
\ee
In particular, one can write the above as 
\be\label{bdyKillingSp.Eq}
\nabla^{3d}_{A}\psi_{-}^{i}=\frac{i}{2}H\gamma_{A}\psi_{-}^{i}+V_{A}\gamma_{3}\psi_{-}^{i}\,,
\ee
for some choice of $H$ and $V_{A}$ and $\psi^{i}_{-}=(\xi_{-},\xi'_{-})$ (see the appendix~\ref{ConventionAndNotation} for the convention). This is the 3-dimensional Killing spinor equation with $H$ and $V_{A}$ being the auxiliary fields in 3-dimensional supergravity.\\
In the following, we will try to find a possible solution for the supergravity background fields $T$, $V$ and $M$ that solves the sets of Killing spinor equation as given in Eqs. \eqref{KillingSpEq1},\eqref{KillingSpEq2} and also at the same time satisfies the condition as given in \eqref{antisyscon} at the boundary. For this we have considered two different kinds of squashing of the four dimensional manifold as already mentioned, namely the one which preserve $SU(2)\times U(1)$ isometry and other with $U(1)\times U(1)$ isometry. The supersymmetric backgrounds for the two cases are separately discussed below.
\subsubsection{Squashed sphere with $SU(2)\times U(1)$ isometry}
In this section, we will consider the supersymmetry on the squashed hemisphere described by the metric~ \eqref{SquashedMetric1}. The squashing was first discussed in~\cite{Cabo-Bizet:2014nia} in the context of supersymmetric localization. The presentation below follows their analysis closely. Our choice of vielbeins are
\bea
&& e^1 = - \dfrac{\sin r}{2} \big(\cos \psi d\theta + \sin \psi \sin \theta d\phi \big)\,,\qquad e^2 =  \dfrac{\sin r}{2} \big(\sin \psi d\theta - \cos \psi \sin \theta d\phi \big)\,, \nn \\
&& e^3 = - \dfrac{\sin r}{2} h(r) \big( d\psi + \cos \theta d\phi \big)\,,\qquad e^4 = dr\,.
\eea
Following \cite{Cabo-Bizet:2014nia}, we choose an ansatz for the Killing spinors that is compatible with the symplectic Majorana reality condition, given in~\eqref{SymMajRealityCondition}. The ansatz is
\be
\xi^{1} = \begin{pmatrix}
	s(r)\\
	0 \\
	i c \frac{h(r)}{s(r)}\sin{r}\\
	0
\end{pmatrix}, \quad
\xi^{2} = \begin{pmatrix}
	0\\
	s(r) \\
	0 \\
	-i c \frac{h(r)}{s(r)}\sin{r}
\end{pmatrix} \quad
\ee 
with $c$ being a real constant. Solving the Killing spinor equation with the above ansatz, we find
\begin{equation}\begin{split}\label{shf}
s(r)&=h(r)\cos{\frac{r}{2}}\,.
\end{split}
\end{equation}
Note that in the limit of vanishing squashing parameter ($\alpha=0$), the above ansatz for the Killing spinor satisfies Killing spinor equation on the round sphere. However for non zero $\alpha$, the above spinors solve the Killing spinor equation provided one turns on appropriate background supergravity fields. These background fields, the antisymmetric matrix $T^{ab}$ and the $SU(2)_{R}$ gauge field $\mathcal{V}_{\mu}$, defined in~\eqref{KillingSpEq1} have the following form,
\be\label{Tmatrix}
T^{ab}\gamma_{ab}= \begin{pmatrix}
	i t^+_3 && i(t^+_1 - i t^+_2) && 0 && 0 \\
	i(t^+_1 + i t^+_2) && -i t^+_3  && 0 && 0 \\
	0 && 0 &&  i t^-_3 && i(t^-_1 - i t^-_2)  \\
	0 && 0 && i(t^-_1 + i t^-_2) && -i t^-_3  \\
\end{pmatrix}\,,
\ee
and
\bea
&& V_a \equiv \begin{pmatrix}
	i v_{3,a} && i (v_{1,a} + i v_{2,a}) \\
	i (v_{1,a} - i v_{2,a}) && - i v_{3,a}
\end{pmatrix}\,,
\eea
with, $\mathcal{V}_\mu = e^a_{\mu} V_a$. The only non zero components of the background fields, $v_{3,3}$, $t^{+}_{3}$, $t^{-}_{3}$ and also $M$ are given in Appendix \eqref{background}.
In presence of the boundary at $r=\pi/2$, we also require to impose the boundary condition as given in \eqref{antisyscon} on the antisymmetric tensor. 
Given the explicit results for the background field components in Eq. \eqref{vt3t3b}, the above condition on the antisymmetric tensor at the boundary translates into the following relation between $t^{+}_{3}$ and $t^{-}_{3}$,
\begin{equation}
\left(t^{+}_{3}-t^{-}_{3}\right)|_{r=\pi/2}=0\Rightarrow c=\pm\left(\frac{1}{2}+\frac{\alpha}{\pi^2}\right).
\end{equation}
In other words, the real constant $c$ is fixed by considering the above boundary condition. Moreover, due to the boundary, we expect only half of the supersymmetry to be present compared to the case with no boundary. We define a projected spinor $\xi^i_{\pm}$ as
\begin{equation}
\xi^{i}_{\pm}=\Pi^{i}_{\pm j}\xi^{j}.
\end{equation}
Imposing the condition that $\xi^{i}_{+}|_{r=\pi/2}=0$ sets the value $c=+\left(\frac{1}{2}+\frac{\alpha}{\pi^2}\right)=\frac{1}{2}h(\frac{\pi}{2})$. The boundary supersymmetry is generated by the Killing spinor $\xi^{i}_{-}\Big|_{r=\frac{\pi}{2}}$.\\
The corresponding Killing vector is given by
\be
K=-4h(\frac{\pi}{2})\frac{\p}{\p\psi}\,.
\ee
The boundary Killing spinor equation is given by~\eqref{bdyKillingSp.Eq} and the supergravity fields are
\be
H=h(\frac{\pi}{2}),\quad V_{A}=-i\delta_{A3}\Big(h(\frac{\pi}{2})-\frac{1}{h(\frac{\pi}{2})}\Big)\,.
\ee
\subsubsection{Squashed sphere with $U(1)\times U(1)$ isometry}
In this section, we will discuss the squashing described by the metric~\eqref{SquashedMetric2}. This particular kind of squashed geometry for S$^{4}$ was considered in~\cite{Hama:2012bg}. The solution of Killing spinor equations and the exact results for the background fields were obtained in~\cite{Hama:2012bg}. The Killing spinors are given by,
\be\label{KillingSp.Hamma}
\xi^{1} = \begin{pmatrix}
	\frac{1}{2}e^{\frac{i}{2}\left(-\theta+\phi+\chi\right)}\sin{\frac{r}{2}}\\
	-\frac{1}{2}e^{\frac{i}{2}\left(\theta+\phi+\chi\right)}\sin{\frac{r}{2}}\\
	\frac{i}{2}e^{\frac{i}{2}\left(-\theta+\phi+\chi\right)}\cos{\frac{r}{2}}\\
	-\frac{i}{2}e^{\frac{i}{2}\left(\theta+\phi+\chi\right)}\cos{\frac{r}{2}}
\end{pmatrix},\quad
\xi^{2} = \begin{pmatrix}
	\frac{1}{2}e^{\frac{i}{2}\left(-\theta-\phi-\chi\right)}\sin{\frac{r}{2}}\\
	\frac{1}{2}e^{\frac{i}{2}\left(\theta-\phi-\chi\right)}\sin{\frac{r}{2}}\\
	-\frac{i}{2}e^{\frac{i}{2}\left(-\theta-\phi-\chi\right)}\cos{\frac{r}{2}}\\
	-\frac{i}{2}e^{\frac{i}{2}\left(\theta-\phi-\chi\right)}\cos{\frac{r}{2}}
\end{pmatrix}\quad
\ee
Given this Killing spinor one can easily obtain the projected Killing spinors $\xi^{i}_{+}$ and $\xi^{i}_{-}$ on the three dimensional boundary. It turns out that $\xi^{i}_{+}$ being proportional to $\left(\cos{\tfrac{r}{2}}-\sin{\tfrac{r}{2}}\right)$ goes to zero at $r=\pi/2$. Moreover, for the Killing spinor equation \eqref{KillingSpEq1} to be satisfied at the boundary with the remaining spinor $\xi^{i}_{-}$, the antisymmetric tensor must obey equation \eqref{antisyscon}. The components of $T_{Bn}$ which are nonzero turn out to be $T_{\phi r}$ and $T_{\chi r}$, given by,
\begin{equation}
T_{\phi r}=\frac{h(r,\theta)\sin{\theta}}{8f(\theta)g(r,\theta)},~~~~T_{\chi r}=-\frac{h(r,\theta)\cos{\theta}}{8f(\theta)g(r,\theta)} \cdot 
\end{equation}
These two components vanish at $r=\pi/2$, since $h(\frac{\pi}{2},\theta)=0$.\\
\ndt The corresponding Killing vector is given by
\be
K=-\frac{1}{\ell}\frac{\p}{\p\phi}-\frac{1}{\tilde\ell}\frac{\p}{\p\chi}\,.
\ee
At the boundary, the Killing spinors~\eqref{KillingSp.Hamma} satisfy the 3-dimensional Killing spinor equation~\eqref{bdyKillingSp.Eq}. The background supergravity fields are given by
\be
H=-\frac{1}{f(\theta)},\quad V_{1}=-\frac{i}{2\cos\theta}\Big(\frac{1}{f(\theta)}-\frac{1}{\ell}\Big),\quad V_{2}=-\frac{i}{2\sin\theta}\Big(\frac{1}{f(\theta)}-\frac{1}{\tilde\ell}\Big),\quad V_{3}=0\,.
\ee
\section{Supersymmetric partition function}
In this section, we will compute the partition function of a supersymmetric theory of interest on the squashed hemisphere using localization. In particular, we will be interested in the supersymmetric theory of $n$ number of chiral multiplets on the boundary of the squashed hemisphere coupled to an abelian vector multiplet having propagating degrees of freedom in bulk. We will further denote by $n_+$ and $n_-$, with $n=n_++n_-$, the number of positively and negatively charged chiral multiplets, respectively. The Lagrangian of the theory on a squashed background is available in the literature~\cite{Hama:2011ea,Hama:2012bg}, however, for the convenience of a reader, we have given the Lagrangian for vector multiplet in the appendix~\ref{SusyLagrangian}. 
\subsection{Squashing preserving $SU(2)\times U(1)$}
We begin with the partition function computation on the squashed hemisphere~\eqref{SquashedMetric1}. The supersymmetric locus is obtained by solving the bosonic equations $\delta\lambda^{i}=0$. To state the solution explicitly, we choose the deformation to be~\eqref{h}.
With this we have a Killing spinor with
\be
s(r)=\cos\frac{r}{2}(1+\frac{2\alpha}{\pi^{2}}\sin^{3}r)\,.
\ee
The solution to the variation $\delta\lambda^{i}=0$ is given by (in perturbative expansion in $\alpha$)
\bea
&&S=s_{0}-\frac{s_{0}\alpha}{\pi^{2}}\sin^{2}\frac{r}{2}(4+5\sin r+4\sin 2r+\sin 3r)+...\nn\\
&&D_{3}=s_{0}+\frac{s_{0}\alpha}{4\pi^{2}}(-8+16\cos r+8\sin r+2\sin 2r+8\sin 3r+5\sin 4r)+...
\eea
In the above $s_{0}$ is constant and all other fields are set to zero.
Evaluating the classical action (see appendix~\ref{SusyLagrangian}) on the localization background we obtain (to order $\alpha^{2}$)\footnote{Note that the partition function depends on $\bar\tau$. This reflects the choice of the Killing spinor. It is possible to find the Killing spinor which gives rise to the classical action on the localization background depending on $\tau$. However, we will not be using the partition function for the future computations, and hence we will not proceed to find such Killing spinor.} 
\be
I=-is_{0}^{2}\Big(\pi-\frac{4\alpha}{\pi}+\frac{12\alpha^{2}}{\pi^{3}}+...\Big)\bar\tau\,,
\ee
where
\be
\bar\tau=\frac{\theta}{2\pi}-\frac{2\pi i}{g^{2}_{YM}}\,.
\ee
The $\alpha$ expansion of the action agrees with the closed form expression of the action given as (we have checked it for a quite a few order)
\be
I=-is_{0}^{2}\frac{\pi}{(1+\frac{2\alpha}{\pi^{2}})^{2}}\bar\tau=-is_{0}^{2}\frac{\pi}{h(\frac{\pi}{2})^{2}}\bar\tau.
\ee 
Thus, the partition function is given by
\be
Z^{{\p\RN{1}}}=\int\, d\sigma\,e^{-i\frac{\pi \sigma^{2}}{h(\frac{\pi}{2})^{2}}\bar\tau}Z^{{\RN{1}}}_{1-\text{loop}}(\sigma,\alpha)\,,
\ee
where the integration contour is chosen along the imaginary direction of $s_{0}$ i.e. $s_{0}=i\sigma$ with $\sigma\in\mathbb R$. 
The one loop determinant for each chiral multiplet is given by~\cite{Hama:2011ea}\footnote{Note that our Killing spinor is normalized differently than~\cite{Hama:2011ea}. In particular, at $r=\frac{\pi}{2}$ it is given by $\bar\xi_{i}\xi^{i}=2h(\frac{\pi}{2})^{2}$. Also, the localization background at $r=\frac{\pi}{2}$ is $S=\frac{s_{0}}{h(\frac{\pi}{2})^{2}}$ and $D_{3}=\frac{s_{0}}{h(\frac{\pi}{2})}$.}
\be
Z^{{\RN{1}}}_{1-\text{loop}}(\sigma,\alpha)=\prod_{n>0}\frac{n+1-q+i\frac{\sigma}{h(\frac{\pi}{2})}}{n-1+q-i\frac{\sigma}{h(\frac{\pi}{2})}}\,.
\ee
Thus, we see that the partition function depends trivially on $h(\frac{\pi}{2})$.
\subsection{Squashing preserving $U(1)\times U(1)$}
Next, we consider the deformation given by~\eqref{SquashedMetric2}. The localization background is obtained by solving the bosonic equations $\delta\lambda^{i}=0$ and is given by~\cite{Hama:2012bg}
\be
D_{1}=-\frac{h(r,\theta)s_{0}}{f(\theta)g(r,\theta)}\sin(\phi+\chi),\quad D_{2}=-\frac{h(r,\theta)s_{0}}{f(\theta)g(r,\theta)}\cos(\phi+\chi),\quad D_{3}=-\frac{s_{0}}{f(\theta)}
\ee
and $s_{0}$ is the constant value of the scalar field $S$. Rest all other fields are zero. The bulk action (see appendix~\ref{SusyLagrangian}) evaluated on the localization background is given by
\be
I_{g}=-\frac{2\pi^{2}}{g^{2}_{YM}}\ell\tilde\ell\,s^{2}_{0}\,.
\ee
Similarly, the boundary action evaluated on the localization background is
\be
I_{\p g}=\frac{i\theta}{2}s^{2}_{0}\,\ell\tilde\ell\,.
\ee
Thus, the complete action on the localization background is given by
\be
I=I_{g}+I_{\p g}=i\ell\tilde\ell s^{2}_{0}\pi\Big(\frac{\theta}{2\pi}+\frac{2\pi i}{g^{2}_{YM}}\Big)=i\ell\tilde\ell\pi\tau s^{2}_{0}\,.
\ee
The partition function is given by
\be\label{SquashedPartitionfn}
Z^{{\p\RN{2}}}=\int d\sigma\,e^{i\ell\tilde\ell\pi\tau \sigma^{2}}Z^{{\RN{2}}}_{1-\text{loop}}\,,
\ee
where the integration contour is chosen along the imaginary direction of $s_{0}$ i.e. $s_{0}=i\sigma$ with $\sigma\in\mathbb R$ and $Z^{{\RN{2}}}_{1-\text{loop}}$ is the one loop contribution from the matter multiplets at the boundary. The one-loop contribution from the $n_+$ and $n_-$ chiral multiplets at the boundary with R-charge $q_+$ and $q_-$, respectively is given in terms of hyperbolic Gamma function~\cite{Hama:2011ea,Imamura:2011wg}
\be
Z^{{\RN{2}}}_{1-\text{loop}}=\Big(\Gamma_{h}(\ell\tilde\ell\sigma+i\omega q_{+};i\omega_{1},i\omega_{2})\Big)^{n_{+}}\Big(\Gamma_{h}(-\ell\tilde\ell\sigma+i\omega q_{-};i\omega_{1},i\omega_{2})\Big)^{n_{-}}\,,
\ee
where $\omega_{1}=b, \omega_{2}=\frac{1}{b}$ and $\omega=\frac{1}{2}(\omega_{1}+\omega_{2})$. The parameter $b$ is the squashing parameter given by $b=\sqrt{\frac{\tilde\ell}{\ell}}$. For our purposes, it is convenient to use the integral representation of the hyperbolic Gamma function. It is given by
\be
\Gamma_{h}(z;w_{1},w_{2})=e^{i\int^{\infty}_{0}\frac{dy}{y}\Big(\frac{z-w}{w_{1}w_{2}y}-\frac{\sin 2y(z-w)}{2\sin w_{1}y\,\sin w_{2}y}\Big)}\,.
\ee
The above integral is well defined for $0<\text{Im}\,z<2\text{Im}\,w$, where $w=\frac{1}{2}(w_{1}+w_{2})$. In the case when there is no squashing, the hyperbolic Gamma function reduces to Jafferis $\ell$-function. The precise relation is 
\be
\Gamma_{h}(z;i,i)=e^{\ell(1+iz)}\,.
\ee
Next, we will use the above partition function to compute the 2-point correlation function of boundary energy-momentum tensor.
\section{$\tau_{R}$ computation}
In this section, we will compute the 2-point function of boundary energy-momentum tensor for various matter degrees of freedom at the boundary interacting with a bulk photon.
The 2-point function of energy momentum tensor in flat space is given by
\bea
<T_{\mu\nu}(x)T_{\rho\sigma}(0)>&=&-\frac{\tau_{R}}{64\pi^2}(\delta_{\mu\nu}\p^2-\p_\mu\p_\nu)(\delta_{\rho\sigma}\p^2-\p_\rho\p_\sigma)\frac{1}{x^2}\nn\\
&&+\frac{\tau_{R}}{64\pi^2}\Big((\delta_{\mu\rho}\p^2-\p_\m\p_\rho)(\delta_{\n\sigma}\p^2-\p_\n\p_\sigma)+(\m\leftrightarrow\n)\Big)\frac{1}{x^2}\,.
\eea
As discussed previously, the quantity of interest, $\tau_R$ can be computed by placing the theory on a squashed sphere. It is given by
\be
\tau_{R}=\frac{2}{\pi^{2}}\text{Re}\frac{\p^2F}{\p b^2}\Big|_{b=1}, \qquad F=-\ln Z^{\p}\,.
\ee
\subsection{Free matter at the boundary}
We begin with the case of the free conformal matter at the boundary i.e. there is no interaction with the bulk photon. This case will be important for our future discussion when we will consider the interaction to be very small.
For a conformal chiral matter, we have $q=\frac{1}{2}$. Thus, the one-loop determinant is given by
\be
Z^{'\text{chiral}}_{1-\text{loop}}=e^{-F}=\Gamma_{h}(\omega(iq);i\omega_{1},i\omega_{2})=\exp\Big(i\int^{\infty}_{0}\frac{dx}{x}\Big[-\frac{\omega(iq)-i\omega}{\omega_{1}\omega_{2}x}+\frac{\sin(2x(iq\omega-\omega))}{2\sinh \omega_{1}x\,\sinh\omega_{2}x}\Big]\Big)\,.
\ee
In the above $\omega_{1}=b,\omega_{2}=\frac{1}{b}$ and $\omega=\frac{1}{2}(\omega_{1}+\omega_{2})$. Thus, the free energy is given as
\be
F=\int^{\infty}_{0}\frac{dx}{x}\Big[-\frac{(b+b^{-1})(q-1)}{2x}+\frac{\sinh(x(b+b^{-1})(q-1))}{2\sinh bx\,\sinh b^{-1}x}\Big]\,.
\ee
Substituting the R-charge $q=\frac{1}{2}$ in the above, we get the free energy 
\be
F=-\int^{\infty}_{0}\frac{dx}{x}\Big[-\frac{(b+b^{-1})}{4x}+\frac{\sinh(\frac{x}{2}(b+b^{-1}))}{2\sinh bx\,\sinh b^{-1}x}\Big]\,.
\ee
Calculating the second derivative w.r.t $b$ and then setting $b=1$, we obtain
\be
\frac{\p^2F}{\p b^2}\Big|_{b=1}=-\int^{\infty}_{0}\frac{dx}{x}\Big(-\frac{1}{2x}-\frac{x}{2}\frac{\cosh x}{\sinh^{2}x}+\frac{x^{2}}{\sinh^{3}x}\Big)=\frac{\pi^{2}}{8}\,.
\ee
Thus, the coefficient $\tau_{R}$ for a free chiral multiplet is
\be
\tau_{R}=\frac{1}{4}\,.
\ee
\subsection{Interacting case: Non chiral theory}
Next, we will consider the partition function of non-chiral matter at the boundary interacting with the bulk photon. It consists of an equal number of chiral multiplets of positively and negatively charged coupled to a single $U(1)$ gauge field. In this case, the squashing dependent partition function is given by\footnote{Here onwards we will consider only the squashing dependent part of the partition function, ignoring the overall factor of $\ell\tilde\ell$.}
\be\label{NonChiralPartitionFn.1}
Z^{\p}=\int\,d\sigma\, e^{i\sigma^{2}\pi\tau}\, (\Gamma_{h}(\sigma+i\omega q_{+};i\omega_{1},i\omega_{2}))^{n}(\Gamma_{h}(-\sigma+i\omega q_{-};i\omega_{1},i\omega_{2}))^{n}\,.
\ee
In the above expression, we have not included the monopole charge $q_{t}$ since it vanishes at the extremum in the present case.

We first consider the case of $n=1$ i.e. of two oppositely charged chiral multiplets.
Using the exponential form of the hyperbolic gamma function, the partition function is given by
\bea
Z^{\p}&=&\int\,d\sigma\, e^{i\sigma^{2}\pi\tau+G(\sigma)}\,,
\eea
where
\be
G(\sigma)=\int^{\infty}_{0}\frac{dy}{y}\Big[\frac{2\omega (q_{f}-1)}{y}\nn\\
+\frac{i\sin(2y(\sigma+i\omega q_{+}-i\omega))+i\sin(2y(-\sigma+i\omega q_{-}-i\omega))}{2\sinh by\,\sinh b^{-1}y}\Big]\,.
\ee
We will further write $q_{\pm}$ as $q_{\pm}=q_{f}\pm q_{g}$.\\
{\bf Large $\tau$ analysis:} We will start with saddle point calculation in the weak coupling limit i.e. $|\tau|>>1$. In this limit, the matter degrees of freedom at the boundary interact weakly with the bulk photon, and as a result, one would expect $\tau_{R}\sim \frac{2}{4}=\frac{1}{2}$ to leading order in large $|\tau|$ expansion. 
We will see this explicitly below. \\  
In the large $|\tau|$ limit, the partition function on S$^{3}$ is extremum at the value $q_{\pm}$ given by~\cite{Gupta:2019qlg} 
\be
q_{g}=0,\quad \text{and}\quad q_{f}=\frac{1}{2}-\frac{\sin\alpha}{\pi|\tau|}-\frac{\pi^{2}-4+(4+(1+2n)\pi^{2})\cos2\alpha}{4\pi^{2}|\tau|^{2}}+\mathcal O(|\tau|^{-3})\,,
\ee
where $\tau=|\tau|e^{i\alpha}$\,.
 We will do saddle point calculation with the above choice of $q_{f}$ and $q_{g}$. We see that $\sigma=0$ is the saddle point, since the function $G(\sigma)$ satisfies 
 \be
 \frac{\p G(\sigma)}{\p \sigma}\Big|_{\sigma=0}=0\,.
 \ee
 Let the function has the following Taylor series expansion about $\sigma=0$
 \be
 G(\sigma)=G_{0}+\frac{1}{2}G_{2}\sigma^{2}+\mathcal O(\sigma^{3})
 \ee
 The value of the above functions are
 \be
 G_{0}=\int^{\infty}_{0}\frac{dy}{y}\Big[\frac{2\omega (q_{f}-1)}{y}\nn\\
-\frac{\sinh(2y\omega (q_{f}-1))}{\sinh by\,\sinh b^{-1}y}\Big]\,,
 \ee
 and
 \be
 G_{2}=\int^{\infty}_{0}dy\Big[\frac{4y\sinh(2y\omega (q_{f}-1))}{\sinh by\,\sinh b^{-1}y}\Big]\,.
 \ee
 Note that both the integrals are entirely convergent since $q_{f}<2$.
 Thus the partition function to a leading order is given by
 \be
 Z^{\p}=e^{G_{0}}\Big(\frac{1}{\sqrt{-i\tau}}+\frac{1}{2}G_{2}\frac{1}{2\pi (-i\tau)^{3/2}}\Big)+\mathcal O(|\tau|^{-5/2})\,.
 \ee
 Thus the real part of free energy is
 \be
 |Z^{\p}|^{2}=e^{-2\text{Re}F_{\p}}=\frac{e^{\bar G_{0}+G_{0}}}{|\tau|}\Big(1+\frac{i}{4\pi|\tau|^{2}}(G_{2}\bar\tau-\bar G_{2}\tau)\Big)+\mathcal O(|\tau|^{-3})\,.
 \ee
Thus the expression for $\tau_{R}$ is
 \be
\tau_{R}=\frac{2}{\pi^{2}}\text{Re}\frac{\p^2F_{\p}}{\p b^2}\Big|_{b=1}=-\frac{1}{\pi^{2}}\frac{\p^2}{\p b^2}(\bar G_{0}+G_{0}+\frac{i}{4\pi|\tau|^{2}}(\bar\tau G_{2}-\tau\bar G_{2}))\Big|_{b=1}
\ee
We will calculate the above one by one 
\be
\frac{\p^2}{\p b^2}(\bar G_{0}+G_{0})\Big|_{b=1}=-\frac{\pi^{2}}{2}-\frac{\pi}{3}\frac{\sin\alpha}{|\tau|}-\frac{\pi}{3}\frac{\sin\alpha}{|\tau|}+\mathcal O(\tau^{-2})=-\frac{\pi^{2}}{2}-\frac{2\pi}{3}\frac{\sin\alpha}{|\tau|}+\mathcal O(\tau^{-2})
\ee
Next we calculate
\be
\frac{\p^2}{\p b^2}(\frac{i}{4\pi|\tau|^{2}}(\bar\tau G_{2}-\tau\bar G_{2}))\Big|_{b=1}=\frac{\text{Im}\tau}{2\pi|\tau|^{2}}\frac{\p^2}{\p b^2}G_{2}\Big|_{b=1}=\frac{\text{Im}\tau}{2\pi|\tau|^{2}}\frac{\pi^{2}}{2}(-8+\pi^{2})=\frac{\pi\text{Im}\tau}{4|\tau|^{2}}(-8+\pi^{2})
\ee
Thus we have
\be\label{nctaurweakcoupling}
\tau_{R}=\frac{1}{2}+\frac{\sin\alpha}{12\pi|\tau|}(32-3\pi^{2})+\mathcal O(\tau^{-2})\,,
\ee
where $\tau=|\tau|e^{i\alpha}$.\\
We can easily extend the above analysis for the case of $n_{\pm}=n$. 
In this case, the partition function is given by
\be
Z^{\p}=\int\,d\sigma\, e^{i\sigma^{2}\pi\tau+\tilde F(\sigma)}\,,
\ee
where
\be
\tilde F(\sigma)=n\int^{\infty}_{0}\frac{dy}{y}\Big[\frac{2\omega (q_{f}-1)}{y}+\frac{i\sin(2y(\sigma+i\omega q_{+}-i\omega))+i\sin(2y(-\sigma+i\omega q_{-}-i\omega))}{2\sinh by\,\sinh b^{-1}y}\Big]
\ee
Since $n$ is an overall factor, one, therefore, has
 \be
\tau_{R}=\frac{n}{2}+\frac{n\sin\alpha}{12\pi|\tau|}(32-3\pi^{2})+\mathcal O(\tau^{-2})\,.
\ee
{\bf Small $\tau$ analysis:} Next, we consider the large coupling expansion, i.e. $|\tau|<<1$, for the coefficient $\tau_{R}$. We will focus here the case of $n_{+}=n_{-}=1$. The partition function is
\be
Z^{\p}=\int\,d\sigma\, e^{i\sigma^{2}\pi\tau}\, \Gamma_{h}(\sigma+i\omega q_{+};i\omega_{1},i\omega_{2})\Gamma_{h}(-\sigma+i\omega q_{-};i\omega_{1},i\omega_{2})\,.
\ee
We will evaluate the above partition function for the value of $q_{f}$ and $q_{g}$ which extremizes the corresponding $S^{3}$ partition function in the limit of $|\tau|<<1$. To the leading order, the extremum occurs at $q_{g}=0$ and
\be\label{ExtremumAtsmallTau}
q_{f}=\frac{1}{3}+\frac{54\sqrt{3}-90\pi+8\sqrt{3}\pi^{2}}{9\Big(8\pi(3\sqrt{3}-2\pi)-27\Big)}\,|\tau|\sin\alpha+\mathcal O(\tau^{2})\,.
\ee
Thus, we have
\be
\tau_{R}=\frac{2}{\pi^{2}}\text{Re}\frac{\p^2F}{\p b^2}\Big|_{b=1}=-\frac{2}{\pi^{2}}\text{Re}\Big(\frac{1}{Z^{\p}}\p^{2}_{b}Z^{\p}\Big|_{b=1}\Big)\,,
\ee
where the RHS is evaluated at~\eqref{ExtremumAtsmallTau}. \\
In the above we have used the relation 
\be
\p_{b}\Gamma_{h}(\pm\sigma+iq_{\pm}\omega;ib,\frac{i}{b})\Big|_{b=1}=0\,.
\ee
Calculating the second derivative w.r.t $b$, we obtain
\be\label{2DerivativebPartitionFn}
\p^{2}_{b}Z^{\p}\Big|_{b=1}=\int\,d\sigma\, e^{i\sigma^{2}\pi\tau}\,e^{\ell(1-i\sigma-q_{-})+\ell(1+i\sigma-q_{+})}\Big(X^{+}(\sigma,q_{+})+X^{-}(\sigma,q_{-})\Big)\,,
\ee
where 
\bea
X^{\pm}(\sigma,q_{\pm})&=&\p^{2}_{b}\Big(i\int^{\infty}_{0}\frac{dx}{x}\Big[\frac{\sin(2x(\pm\sigma+iq_{\pm}\omega-i\omega))}{2\sinh bx\,\sinh\frac{x}{b}}-\frac{\sigma+i\omega q_{\pm}-i\omega}{x}\Big]\Big)\Big|_{b=1}\,,\nn\\
&=&\int^{\infty}_{0}dx\,\Big[(q_{\pm}-1)\Big(\frac{1}{x^{2}}-\frac{\cosh(2x(1-q_{\pm}\mp i\sigma))}{\sinh^{2}x}\Big)-\frac{\sinh (2x)-2x}{2\sinh^{4}x}\sinh(2x(1-q_{\pm}+\pm 2ix\sigma))\Big]\nn\\
\eea
In the above we have used the relation
\be
\Gamma_{h}(z;i,i)=e^{\ell(1+iz)}\,,
\ee
where $\ell(z)$ is the Jafferis $\ell$-function.
To evaluate~\eqref{2DerivativebPartitionFn} at strong coupling, we go to the dual frame where the computation reduces to the weak coupling computation. The dual frame is obtained by using the Fourier transform~\cite{Gupta:2019qlg}
\be
\int d\sigma\,e^{\ell(1-q_{+}+i\sigma)+\ell(1-q_{-}-i\sigma)+2\pi i\kappa\sigma}=e^{\ell(1-q_{+}-q_{-})+\ell(\frac{q_{+}+q_{-}}{2}+i\k)+\ell(\frac{q_{+}+q_{-}}{2}-i\k)+\pi\k(q_{+}-q_{-})}\,.
\ee
Using the above identity, we obtain
\bea
\p^{2}_{b}Z^{\p}\Big|_{b=1}&=&
\int\,d\sigma\, \,d\k\,e^{-2\pi i\k\sigma}e^{i\sigma^{2}\pi\tau}\,e^{\ell(1-q_{+}-q_{-})+\ell(i\k+\frac{q_{-}+q_{+}}{2})+\ell(-i\k+\frac{q_{-}+q_{+}}{2})}e^{\pi \k(q_{+}-q_{-})}\nn\\
&&\times\Big(X^{+}(\sigma,q_{+})+X^{-}(\sigma,q_{-})\Big)\,,\nn\\
&=&\int\,d\k\,e^{-\frac{i\pi \k^{2}}{\tau}}\,e^{\ell(1-q_{+}-q_{-})+\ell(i\k+\frac{q_{-}+q_{+}}{2})+\ell(-i\k+\frac{q_{-}+q_{+}}{2})}e^{\pi \k(q_{+}-q_{-})}\nn\\
&&\times\int_{\text{Im}\tilde\sigma=\frac{\k\text{Im}\tau}{|\tau|^{2}}}\,d\tilde\sigma\,e^{i\tilde\sigma^{2}\pi\tau}\Big(X^{+}(\tilde\sigma+\frac{\k}{\tau},q_{+})+X^{-}(\tilde\sigma+\frac{\k}{\tau},q_{-})\Big)\,,\nn\\
&=&\int\,d\k\,e^{-\frac{i\pi \k^{2}}{\tau}}\,e^{\ell(1-q_{+}-q_{-})+\ell(i\k+\frac{q_{-}+q_{+}}{2})+\ell(-i\k+\frac{q_{-}+q_{+}}{2})}e^{\pi \k(q_{+}-q_{-})}\nn\\
&&\times\int^{\infty}_{-\infty}\,d\tilde\sigma\,e^{i\tilde\sigma^{2}\pi\tau}\Big(X^{+}(\tilde\sigma+\frac{p}{\tau},q_{+})+X^{-}(\tilde\sigma+\frac{p}{\tau},q_{-})\Big)\,,\nn\\
&=&\int\,d\k\,e^{-\frac{i\pi \k^{2}}{\tau}}\,e^{\ell(1-2q_{f})+\ell(i\k+q_{f})+\ell(-i\k+q_{f})}\nn\\
&&\times\int^{\infty}_{0}\,\frac{dx}{\sqrt{-i\tau}}e^{-\frac{ix^{2}}{\pi\tau}}\Big[e^{\frac{ix^{2}}{\pi\tau}}\frac{2(q_{f}-1)}{x^{2}}
-\frac{2(q_{f}-1)}{\sinh^{2}x}\cos\frac{2px}{\tau}\cosh(2x(q_{f}-1))\nn\\
&&-\frac{\sinh(2x)-2x}{\sinh^{4}x}\cos\frac{2px}{\tau}\sinh(2x(1-q_{f}))\Big]\,.\nn\\
\eea 
In the above we have deformed the contour back to the real axis and also substituted $q_{\pm}=q_{f}$, since $q_{g}=0$. In the limit $|\tau|\sim 0$, we evaluate the above integral in the saddle point approximation. In this approximation, we get
\bea
\frac{1}{Z^{\p}}\p^{2}_{b}Z^{\p}\Big|_{b=1}&=&\int^{\infty}_{0}\,dx\Big[\frac{2(q_{f}-1)}{x^{2}}
+\frac{1}{2}e^{-2\ell(\frac{1}{3})+\ell(i\frac{x}{\pi}+\frac{1}{3})+\ell(-i\frac{x}{\pi}+\frac{1}{3})}(2+|\tau|f_{1}(x)+|\tau|f_{1}(-x)-2|\tau|f_{1}(0))\times\nn\\
&&\times\Big(-\frac{2(q_{f}-1)}{\sinh^{2}x}\cosh(2x(q_{f}-1))-\frac{\sinh(2x)-2x}{\sinh^{4}x}\sinh(2x(1-q_{f}))\Big)\Big]+\mathcal O(|\tau|^{2})\,,
\eea
where the function $f_{1}(x)$ is given in the appendix~\ref{Functions}. Thus, we have
\bea\label{NonChiralLargeCoupling}
\tau_{R}&=&-\frac{2}{\pi^{2}}\text{Re}\int^{\infty}_{0}\,dx\Big[\frac{2(q_{f}-1)}{x^{2}}
+\frac{1}{2}e^{-2\ell(\frac{1}{3})+\ell(i\frac{x}{\pi}+\frac{1}{3})+\ell(-i\frac{x}{\pi}+\frac{1}{3})}(2+|\tau|f_{1}(x)+|\tau|f_{1}(-x)-2|\tau|f_{1}(0))\times\nn\\
&&\times\Big(-\frac{2(q_{f}-1)}{\sinh^{2}x}\cosh(2x(q_{f}-1))-\frac{\sinh(2x)-2x}{\sinh^{4}x}\sinh(2x(1-q_{f}))\Big)\Big]+\mathcal O(|\tau|^{2})\,,
\eea
with the R-charge $q_{f}$ to the first order in $|\tau|$ given in \eqref{ExtremumAtsmallTau}. We could not evaluate the above integral explicitly; however, numerical evaluation is possible for the various values of the angle $\alpha$. In particular, for $\alpha=\frac{\pi}{2}$, we obtain
\be
\tau_{R}=0.545-0.07035 |\tau|+\mathcal O(|\tau|^{2})\,.
\ee
\begin{figure}
\begin{center}
\includegraphics[width=3in]{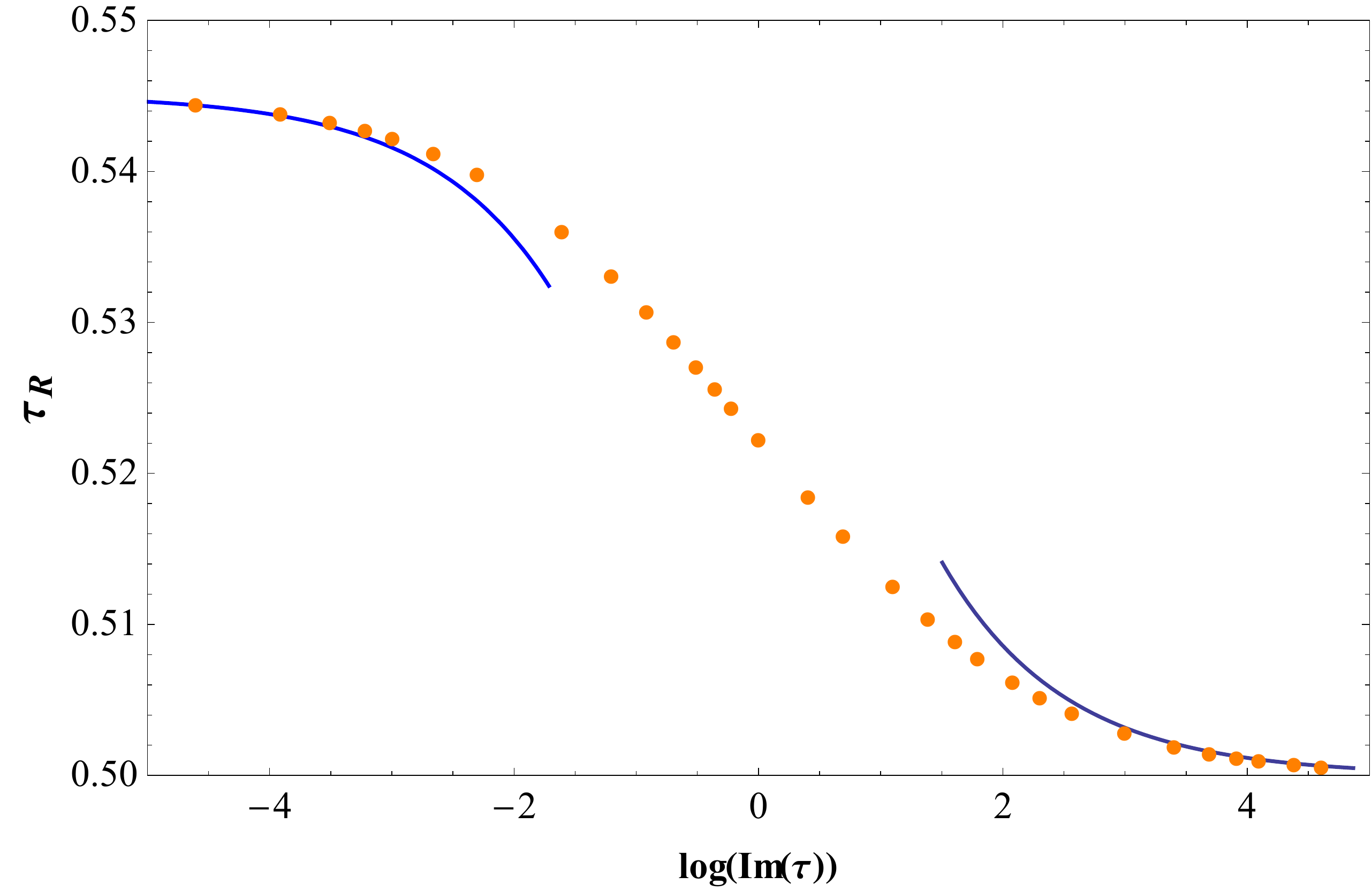}
\end{center}
\caption{The plot of $\tau_{R}$ vs. $\text{Im}\tau$ for the $n_{+}=n_{-}=1$ theory. The dots represent the numerical computations and the solid lines are the saddle point approximation.\label{fig:NonChiralPlot} }
\end{figure}
The full behaviour of $\tau_{R}$ as a function of $\text{Im}\tau$ can be seen in the figure \ref{fig:NonChiralPlot}. 
\\
{\bf Large n-analysis:} We can also find the expression for $\tau_R$ in the large $n$-limit. In this case, we consider the partition function given by
\be
Z^{\p}=\int\,d\sigma\, e^{i\sigma^{2}\pi\tau}\, (\Gamma_{h}(\sigma+i\omega q_{+};i\omega_{1},i\omega_{2}))^{n}(\Gamma_{h}(-\sigma+i\omega q_{-};i\omega_{1},i\omega_{2}))^{n}\,.
\ee
The computation of $\tau_{R}$ in $\frac{1}{n}$-expansion proceeds in a similar manner as in previous cases and we will not repeat here. To compute $\tau_{R}$, we need to know the R-charge which maximize the free energy on $S^{3}$. In the $\frac{1}{n}$-expansion it is given by~\cite{Gupta:2019qlg}
\be
q_{f}=\frac{1}{2}-\frac{2}{\pi^{2}n}+\mathcal O(n^{-2}),\quad \text{and}\quad q_{g}=0\,.
\ee 
Using the above expression for the R-charge, the explicit computation then gives
\be
\tau_{R}=\frac{n}{2}-2\Big(1-\frac{8}{3\pi^{2}}\Big)+\mathcal O(n^{-1})\,.
\ee

\subsection{Interacting case: Chiral theory}
Next, we consider the chiral theory with a positively charged matter field at the boundary i.e. $n_{+}=1, n_{-}=0$. The squashing dependent partition function is given by
\be
Z_{\text{chiral}}^{\p}=\int\,d\sigma\, e^{i\sigma^{2}\pi\tau}\, \Gamma_{h}(\sigma+i\omega q_{g};i\omega_{1},i\omega_{2})e^{2\pi \omega q_{t}\sigma}\,,
\ee
where $q_{t}$ is the monopole charge.
Note that since $\p_{b}\omega\Big|_{b=1}=0$, we have a vanishing one point function of energy momentum tensor
\be
\p_{b}Z_{\text{chiral}}^{\p}\Big|_{b=1}=0\,.
\ee
For the computation of $\tau_{R}$, we need the second derivative of the partition function with respect to the squashing parameter and is given by
\be\label{SecondDerivativeChiralcase.1}
\p^{2}_{b}Z_{\text{chiral}}^{\p}\Big|_{b=1}=\int\, d\sigma\,e^{i\pi\tau\sigma^{2}+\ell(1-q_{g}+i\sigma)+2\pi q_{t}\sigma}X(\sigma)\,,
\ee
where
\bea\label{DefinitionXk}
X(\sigma)&=&2\pi\sigma q_{t}+\int^{\infty}_{0}dx\,\Big[\frac{q_{g}-1}{x^{2}}+\frac{(1-q_{g})\cosh (2x(q_{g}-1)-2ix\sigma)}{\sinh^{2}x}\nn\\
&&-\frac{1}{2\sinh^{4}x}(\sinh (2x)-2x)\sinh(2x(1-q_{g})+2ix\sigma)\Big]
\eea
We first calculate the $\tau_{R}$ in the weak coupling limit i.e. $|\tau|\rightarrow\infty$. The superconformal R-symmetry is determined by the value of $q_{g}$ and $q_{t}$ that extremize the partition function on S$^{3}$. In the saddle point approximation, these are given by~\cite{Gupta:2019qlg}
\be
q_{g}=\frac{1}{2}-\frac{\text{Im}\tau}{\pi|\tau|^{2}}+\mathcal O(|\tau|^{-2}),\quad q_{t}=-\frac{\text{Re}\tau}{4|\tau|^{2}}+\mathcal O(|\tau|^{-2})\,.
\ee
The computation of $\tau_R$ proceeds similarly as in the non-chiral case and we will not repeat here.
The saddle point approximation in the limit $|\tau|\rightarrow\infty$ gives
\be
\tau_{R}=\frac{1}{4}+\frac{\sin\alpha}{24\pi|\tau|}(32-3\pi^{2})+\mathcal O(\tau^{-2})\,.
\ee
Note that to the order $\mathcal O(\frac{1}{|\tau|})$, the result of $\tau_{R}$ is exactly half of the non chiral case~\eqref{nctaurweakcoupling}. 
\begin{figure}
\begin{center}
\includegraphics[width=3in]{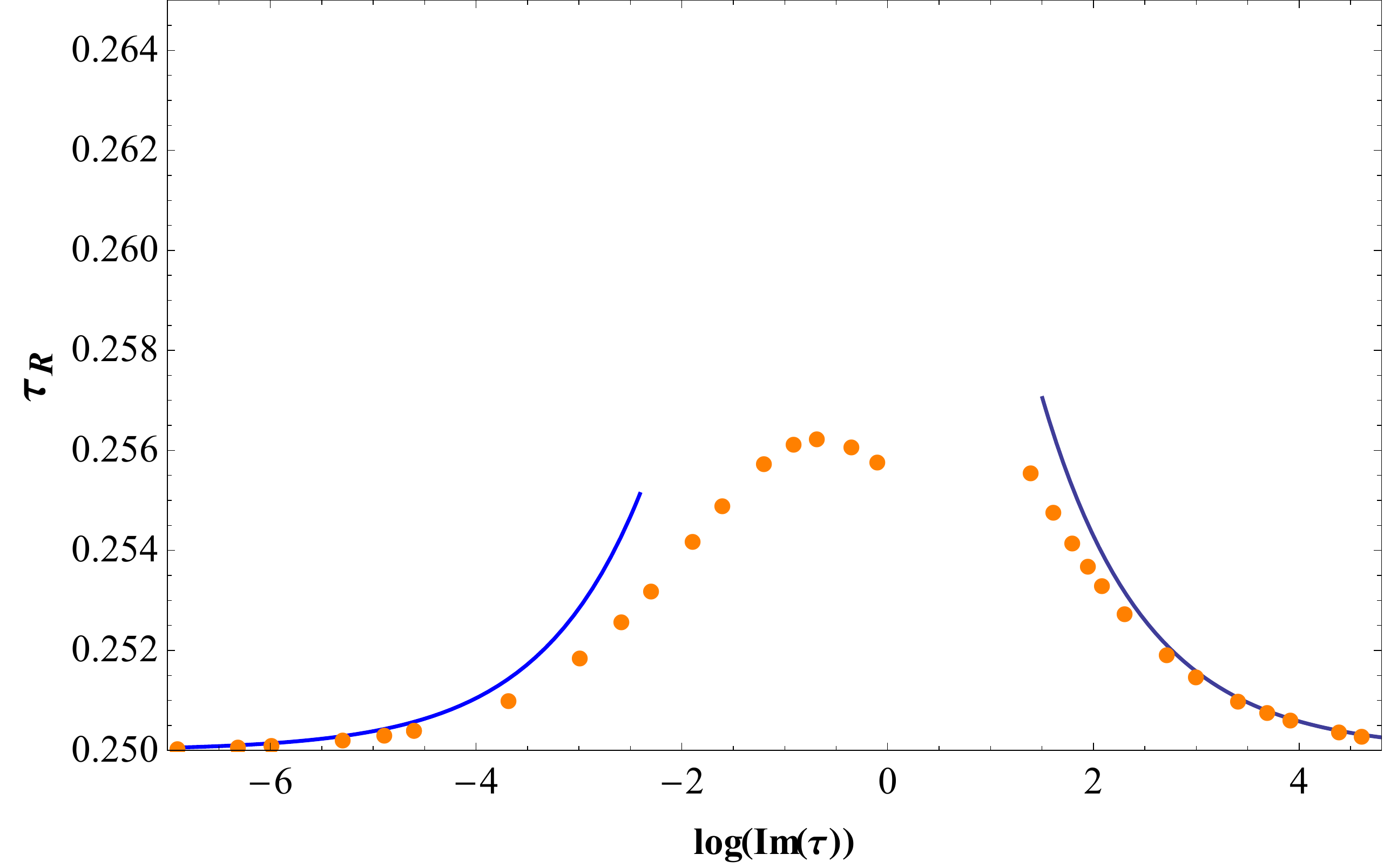}
\end{center}
\caption{The plot of $\tau_{R}$ vs. $\text{Im}\tau$ for the $n_{+}=1, n_{-}=0$ theory. The dots represent the numerical computations and the solid lines are the saddle point approximation.\label{fig:ChiralPlot} }
\end{figure}
For the computation of $\tau_{R}$ at the strong coupling $|\tau|<1$, we will use the following identity~\cite{Gupta:2019qlg}
\be
\int d\sigma\,e^{\frac{i\pi}{2}\sigma^{2}+2\pi i\sigma \k+\ell(1-q_{g}+i\sigma)}=e^{-2\pi i\k^{2}+\ell(\frac{q_{g}+1}{2}+i\k)-\frac{3\pi i}{8}(q_{g}+2i\k-\frac{1}{3})^{2}+\frac{i\pi}{12}}\,.
\ee
Using the above identity, the derivative of the partition function can be written as
\bea
\p^{2}_{b}Z_{\text{chiral}}^{\p}\Big|_{b=1}&=&\int\, d\k\,dp\,e^{-2\pi ip\k}e^{2\pi q_{t}\k}e^{i\pi(\tau-\frac{1}{2})\k^{2}}e^{-2\pi i p^{2}}e^{\ell(\frac{\tilde q+1}{2})-\frac{3\pi i}{8}(\tilde q-\frac{1}{3})^{2}+\frac{i\pi}{12}}X(\k)\,,
\eea
where the function $X(\k)$ is given in~\eqref{DefinitionXk}.
After some simplifications and performing the integration over $\k$, the integral on the RHS can be written as
\bea\label{SecondDerivativeChiralcase.2}
\p^{2}_{b}Z_{\text{chiral}}^{\p}\Big|_{b=1}=e^{\frac{\pi i}{2} q^{2}_{t}-\frac{3\pi i}{8}(q_{g}-\frac{1}{3})^{2}-\frac{3\pi i}{2}q_{t}(q_{g}-\frac{1}{3})+\frac{i\pi}{12}}\int\,dp\,e^{-\frac{\pi i}{2} p^{2}}e^{\ell(\frac{ q_{g}+1}{2}+q_{t}+ik)+2\pi p(\frac{3q_{g}-1}{4}-\frac{q_{t}}{2})}\tilde Y(p)\,.\nn\\
\eea
with
\bea
\tilde Y(k)&=&2\pi q_{t}\frac{e^{i\frac{\pi}{4}}}{\sqrt{-i(\tau-\frac{1}{2})^{3}}}e^{-\frac{i\pi}{\tau-\frac{1}{2}}k^{2}}k+\int^{\infty}_{-\infty}dx\,\frac{1}{2\sqrt{-i(\tau-\frac{1}{2})}}\Big[\frac{q_{g}-1}{x^{2}}e^{-\frac{i\pi}{\tau-\frac{1}{2}}k^{2}}\nn\\
&&+\frac{(1-q_{g})}{\sinh^{2}x}e^{2x(q_{g}-1)}e^{-\frac{i}{\pi(\tau-\frac{1}{2})}(\pi k+x)^{2}}
+\frac{(\sinh (2x)-2x)}{2\sinh^{4}x}e^{2x(q_{g}-1)}e^{-\frac{i}{\pi(\tau-\frac{1}{2})}(\pi k+x)^{2}}\Big]\,.\nn\\
\eea
Next, we compute the integral~\eqref{SecondDerivativeChiralcase.2} in saddle point approximation when the coupling $\tau\sim \frac{1}{2}$. Let us define a complex coupling $\tau'$ by
\be
\tau'=-\frac{1}{2}-\frac{1}{\tau-\frac{1}{2}}\,.
\ee 
In the limit $|\tau'|>>1$, the superconformal R-symmetry is given by
\be
q_{g}=\frac{1}{4}+\mathcal O(|\tau'|^{-2}),\quad q_{g}+2q_{t}=\frac{2\sin\alpha'}{\pi|\tau'|}+\frac{1}{|\tau'|^{2}}\Big(\frac{3}{4}-\frac{3}{\pi^{2}}+(\frac{5}{4}+\frac{3}{\pi^{2}})\cos2\alpha'\Big)+\mathcal O(|\tau'|^{-3})\,,
\ee
where $\tau'=|\tau'|e^{i\alpha'}$. Evaluating the integral~\eqref{SecondDerivativeChiralcase.2} in the saddle point approximation, using the above R-charge, we find to the first subleading order 
\be\label{TauRChiral}
\tau_{R}=\frac{1}{4}-\frac{2}{\pi^{2}}\text{Re}\frac{1}{\tau'}\int_{-\infty}^{\infty}dx\,g(x,\alpha')+\mathcal O(\tau'^{-2})\,,
\ee
where the explicit form the function $g(x,\alpha')$ is given in the appendix~\ref{Functions}. For $\alpha'=\frac{\pi}{2}$, we have 
\be
\tau_{R}=\frac{1}{4}+\frac{0.0570}{|\tau'|}+\mathcal O(|\tau'|^{-2})\,.
\ee
The full behaviour of $\tau_{R}$ as a function of $\text{Im}\tau$ can be seen in the figure \ref{fig:ChiralPlot}. 
\section{Conclusion}
Our original motivation for the present work is to compute the correlation function of the energy-momentum tensor for the 4-dimensional bCFT. The most general form of the 2-point function of the energy-momentum tensor requiring the conservation law and conformal invariance is known ~\cite{McAvity:1993ue, McAvity:1995zd, Herzog:2017xha}. It is given in terms of 3 arbitrary functions that encode the dynamics of the bCFT. Our goal was to compute these coefficients in a supersymmetric bCFT.
In this direction, we computed the partition function of the mixed dimensional supersymmetric dimensional QED on squashed hemispheres. We considered two different squashings that preserve either $SU(2)\times U(1)$ or $U(1)\times U(1)$ isometry of the original sphere. The boundary free energy depends on the deformation parameter. In the case of the squashing that preserves $SU(2)\times U(1)$ isometry of the original sphere, the free energy is trivial as a function of the deformation. In contrast, the boundary-free energy depends on the squashing parameter in the case of the squashing that preserves $U(1)\times U(1)$ isometry. We then computed the coefficient $\tau_R$ by differentiating the free energy twice with respect to the squashing parameter. On important feature of  $\tau_R$ is that it depends on the bulk marginal coupling $\tau$. We then find the behaviour of $\tau_R$ as we change the coupling. For the non-chiral case, we find that the coefficient $\tau_R$ decreases from strong coupling to weak coupling. In contrast, the computation in the chiral case reveals that it first increases and then decreases as we change the coupling. It would be interesting to understand the physical reason behind it. 

As a future direction, it would be interesting to extend the above computation for the case when the bulk involves non-abelian gauge fields, in particular for the $\mathcal N=4$ SYM. See~\cite{VanRaamsdonk:2020djx} for a recent work in this direction.  
Another direction would be to compute the anomalous contributions to the trace of the energy-momentum tensor. These contributions depend on the extrinsic curvature of the boundary.  For the computation of these contributions, we need to consider the manifold where the boundary has non zero extrinsic curvature. The squashed hemisphere has a vanishing extrinsic curvature. It would be interesting to see if it is possible to find a squashing deformation that preserves some supersymmetry so that the localization computation can be done, and also has non zero extrinsic curvature. For example, one way to obtain non zero extrinsic curvature is to start with a 4-dimensional sphere and put the boundary at $r\neq \frac{\pi}{2}$. 

\section*{Acknowledgments}
We thank Imtak Jeon for useful discussion. AR thanks Arnab Kundu, Dharmesh Jain and Rudranil Basu for various discussions on supersymmetric field theories. He is also grateful for the hospitality extended by IIT, Ropar during Jan., 2020 where parts of the project were discussed. He is supported by ``Senior Research Fellowship'' under Government of India. 
\appendix
\section{Charge Conjugation and gamma matrices}\label{ConventionAndNotation}
In Euclidean signature, the Clifford algebra is given by
\be
\{\gamma^a, \gamma^b \} = 2\, \delta^{ab} \ ,
\ee
where the gamma matrices are Hermitian $(\gamma^a)^\dagger = \gamma^a$.
Our convention for the gamma matrices is the same that Narain et al.\ \cite{Gava:2016oep}  use
\be
\gamma^a = \left( \begin{array}{cc}
0 & -\i \sigma^a \\
\i \sigma^a & 0 
\end{array}
\right) \ ; \; \; \;
\gamma^4 =  \left( \begin{array}{cc}
0 & 1 \\
1 & 0 
\end{array}
\right) \ , \; \; \;
\gamma^5 =  \gamma^1 \gamma^2 \gamma^3 \gamma^4 = \left( \begin{array}{cc}
1 & 0 \\
0 & -1 
\end{array}
\right) \ .
\ee

The 4-dimensional charge conjugation matrix is $C = \gamma_1 \gamma_3 = {\rm diag}(-\i \sigma_2, -\i \sigma_2)$.  We choose $C$ to define symplectic Majorana fermions
\be
\bar \xi_i \equiv (\xi_i)^\dagger = \epsilon_{ij} (\xi^j)^T C \ .
\ee
In particular, $\xi_1^\dagger = \xi^{2T} C$ and $\xi_2^\dagger = -\xi^{1T} C$.  Furthermore, $\xi^2 = C \xi_1^*$.  
The matrix $C$ satisfies the further conjugation properties
\be
\gamma_a^* = \gamma_a^T = C \gamma_a C^{-1} \ , \; \; \; C^\dagger = C^T = C^{-1} = -C \ , \; \; \; C^* = C \ .
\ee

Let us denote the component of the projector as
\be
\Pi^{1}_{\pm1}=\pi_{\pm},\qquad \Pi^{2}_{\pm2}=\pi_{\mp}
\ee
where
\be
\pi_{\pm}=\frac{1}{2}(1\pm i\gamma_{5}\gamma_{n})\,.
\ee
These satisfy
\be
C\pi_{\pm}=\pi^{T}_{\mp}C,\qquad \gamma_{A}\pi_{\pm}=\pi_{\pm}\gamma_{A}\,.
=\psi_{\mp}\gamma_{n}
\ee
Next, we define the relation between the 4 dimensional gamma matrices, $\gamma_{A}$, and 3 dimensional gamma matrices, $\Gamma_{A}$. It is given by
\be
\Gamma_{A}=\pi_{+}\gamma_{A}
\ee
With this we see that
\be
\Gamma^{T}_{A}=\gamma^{T}\pi^{T}_{+}=C\gamma_{A}C^{-1}
\ee
The above gamma matrices satisfy 3 dimensional transpose relation with the charge conjugation matrix
\be
\tilde C=C\gamma_{n},\quad\Rightarrow\quad \Gamma^{T}_{A}=-\tilde C\Gamma_{A}\tilde C^{-1}\,.
\ee
Given a spinor $\psi^{i}$, the projected spinor is given by $\psi^{i}_{\pm}=\Pi^{i}_{\pm j}\psi^{j}$.
Furthermore, we define
\be
\psi^{1}_{\pm}=\psi_{\pm},\quad \psi^{2}_{\pm}=\gamma_{n}\psi'_{\pm},\qquad \text{where}\quad \pi_{\pm}\psi'_{\pm}=\psi'_{\pm}\,.
\ee
 We find the reality condition
\be
\psi^{\dagger}_{\pm}=\psi'^{T}_{\pm}\tilde C,\quad \psi'^{\dagger}_{\pm}=-\psi^{T}_{\pm}\tilde C
\ee
Thus, the pairs $(\psi_{\pm},\psi'_{\pm})$ satisfy the condition of 3-dimensional sympletctic Majorana condition.
\section{Projector and Killing Spinor}\label{ProjectorAndKilling}
In this section, we discuss in details the supersymmetry on a squashed hemisphere. Now, the presence of boundary breaks half of the supersymmetry transformation, i.e. the boundary in our case preserves only 4 out of 8 supercharges. For the squashing deformation to preserve the supersymmetry on the hemisphere, the background needs to satisfy certain conditions. We begin with the projector, which was discussed in the case of supersymmetric field theory on the hemisphere~\cite{Gupta:2019qlg}. It is given by
\be
\Pi^{i}_{\pm\,j}=\frac{1}{2}(\delta^{i}_{j}\pm i\tau^{i}_{3\,j}\gamma_{5}\gamma_{n})\,.
\ee
Here, $\gamma_{n}$ is the flat space gamma matrix corresponding to the direction perpendicular to the boundary. The projector satisfies the following conditions
\be
\gamma_{A}\Pi^{i}_{\pm\,j}=\Pi^{i}_{\pm\,j}\gamma_{A},\qquad \gamma_{n}\Pi^{i}_{\pm\,j}=\Pi^{i}_{\mp\,j}\gamma_{n}\,,
\ee
where $A$ is the boundary index.\\
Given a Killing spinor $\xi^{i}$ on the squashed sphere, we define the following
\be
\xi^{i}_{\pm}=\Pi^{i}_{\pm\,j}\xi^{j}\,.
\ee
We require the projector so that half of the supersymmetry transformation parameter vanishes at the boundary. In other words, we require that if $\xi$ be a Killing spinor, then at $r=\frac{\pi}{2}$, $\Pi^{i}_{+\,j}\xi^{j}\Big|_{r=\frac{\pi}{2}}=0$ and $\xi^{i}_{-}\Big|_{r=\frac{\pi}{2}}$ generates the supersymmetry at the boundary.\\
We start with the Killing spinor equation on the squashed S$^{4}$ 
\be
\cD_{\m}\xi^{i}+\gamma^{ab}T_{ab}\gamma_{\m}\xi=\frac{1}{4}\gamma_{\m}(\gamma^{\n}\cD_{\n}\xi^{i})\,,
\ee 
where the auxiliary field $T_{ab}$ is non zero for the squashing.
Let $A,B,C..$ be the boundary flat indices and $n$ be the normal direction flat index. Then
\be
\cD_{A}\xi^{i}+\gamma^{ab}T_{ab}\gamma_{A}\xi=\frac{1}{4}\gamma_{A}(\gamma^{\n}\cD_{\n}\xi^{i})\,,
\ee
which can be written in terms of 3-dimensional covariant derivative as
\bea\label{To3dKillingSp.}
&&\nabla^{3d}_{A}\xi^{i}+\mathcal V^{i}_{A\,j}\xi^{j}+\frac{1}{2}\omega^{Bn}_{A}\gamma_{Bn}\xi^{i}+(T_{BC}\gamma_{BC}+2T_{Bn}\gamma_{Bn})\gamma_{A}\xi^{i}\nn\\
&&=\frac{1}{4}\gamma_{A}(\gamma^{B}\nabla^{3d}_{B}\xi^{i}+\g^{B}\mathcal V^{i}_{B\,j}\xi^{j}+\gamma^{n}\cD_{n}\xi^{i}+\frac{1}{2}\gamma^{B}\omega^{Cn}_{B}\gamma_{Cn}\xi^{i})\,.\nn\\
\eea
Normal direction Killing spinor equation is
\be
\cD_{n}\xi^{i}+\gamma^{ab}T_{ab}\gamma_{n}\xi=\frac{1}{4}\gamma_{n}(\gamma^{\n}\cD_{\n}\xi^{i})\,,
\ee
which can be further written as
\be
\gamma^{n}\cD_{n}\xi^{i}+\gamma^{n}\gamma^{ab}T_{ab}\gamma_{n}\xi=\frac{1}{4}(\gamma^{A}\nabla^{3d}_{A}\xi^{i}+\g^{B}\mathcal V^{i}_{B\,j}\xi^{j}+\frac{1}{2}\gamma^{B}\omega^{Cn}_{B}\gamma_{Cn}\xi^{i}+\gamma^{n}\cD_{n}\xi^{i})\,.
\ee
Solving the above we get
\be
\gamma^{n}\cD_{n}\xi^{i}=\frac{1}{3}(\gamma^{A}\nabla^{3d}_{A}\xi^{i}+\g^{B}\mathcal V^{i}_{B\,j}\xi^{j}+\frac{1}{2}\gamma^{B}\omega^{Cn}_{B}\gamma_{Cn}\xi^{i}-4\gamma^{n}\gamma^{ab}T_{ab}\gamma_{n}\xi^{i})\,.
\ee
Substituting the above in \eqref{To3dKillingSp.}, we get
\bea
\nabla^{3d}_{A}\xi^{i}+\mathcal V^{i}_{A\,j}\xi^{j}+\frac{1}{2}\omega^{Bn}_{A}\gamma_{Bn}\xi^{i}+(T_{BC}\gamma_{BC}+2T_{Bn}\gamma_{Bn})\gamma_{A}\xi^{i}&=&\frac{1}{4}\gamma_{A}\Big(\frac{4}{3}\gamma^{B}\nabla^{3d}_{B}\xi^{i}+\frac{4}{3}\g^{B}\mathcal V^{i}_{B\,j}\xi^{j}\nn\\
&&-4\gamma^{n}\gamma^{ab}T_{ab}\gamma_{n}\xi\+\frac{2}{3}\gamma^{B}\omega^{Cn}_{B}\gamma_{Cn}\xi^{i}\Big)\,.\nn\\
\eea
Simplifying further, we obtain
\bea
&&\nabla^{3d}_{A}\xi^{i}+\mathcal V^{i}_{A\,j}\xi^{j}+\frac{1}{2}\omega^{Bn}_{A}\gamma_{Bn}\xi^{i}+T_{BC}(\gamma_{BC}\gamma_{A}+\frac{1}{3}\gamma_{A}\gamma_{BC})\xi^{i}-2T_{Bn}(\gamma_{B}\gamma_{A}\gamma_{n}+\frac{1}{3}\gamma_{A}\gamma_{B}\gamma_{n})\xi^{i}\nn\\
&&\qquad\qquad\qquad=\frac{1}{3}\gamma_{A}\gamma^{B}\cD^{3d}_{B}\xi^{i}+\frac{1}{3}\gamma_{A}\g^{B}\mathcal V^{i}_{B\,j}\xi^{j}+\frac{1}{6}\gamma_{A}\gamma^{B}\omega^{Cn}_{A}\gamma_{Cn}\xi^{i}\,.
\eea
In the above $\nabla^{3d}_{A}$ is the 3-dimensional spacetime covariant derivative. Now, we apply the projector $\Pi^{i}_{\pm j}$ and evaluate the above at $r=\frac{\pi}{2}$. We get with $\Pi^{i}_{- j}$
\bea\label{NegativeProj.}
&&\cD^{3d}_{A}\xi_{-}^{i}+ V^{i}_{A\,j}\xi_{+}^{j}+\frac{1}{2}\omega^{Bn}_{A}\gamma_{Bn}\xi_{+}^{i}+T_{BC}(\gamma_{BC}\gamma_{A}+\frac{1}{3}\gamma_{A}\gamma_{BC})\xi_{-}^{i}-2T_{Bn}(\gamma_{B}\gamma_{A}\gamma_{n}+\frac{1}{3}\gamma_{A}\gamma_{B}\gamma_{n})\xi_{+}^{i}\nn\\
&&\qquad\qquad\qquad=\frac{1}{3}\gamma_{A}\gamma^{B}\cD^{3d}_{B}\xi_{-}^{i}+\frac{1}{3}\gamma_{A}\g^{B}V^{i}_{B\,j}\xi_{+}^{j}+\frac{1}{6}\gamma_{A}\gamma^{B}\omega^{Cn}_{A}\gamma_{Cn}\xi_{+}^{i}\,,
\eea
and with $\Pi^{i}_{+j}$
\bea\label{PositiveProj.}
&&\cD^{3d}_{A}\xi_{+}^{i}+ V^{i}_{A\,j}\xi_{-}^{j}+\frac{1}{2}\omega^{Bn}_{A}\gamma_{Bn}\xi_{-}^{i}+T_{BC}(\gamma_{BC}\gamma_{A}+\frac{1}{3}\gamma_{A}\gamma_{BC})\xi_{+}^{i}-2T_{Bn}(\gamma_{B}\gamma_{A}\gamma_{n}+\frac{1}{3}\gamma_{A}\gamma_{B}\gamma_{n})\xi_{-}^{i}\nn\\
&&\qquad\qquad\qquad=\frac{1}{3}\gamma_{A}\gamma^{B}\cD^{3d}_{B}\xi_{+}^{i}+\frac{1}{3}\gamma_{A}\g^{B}V^{i}_{B\,j}\xi_{-}^{j}+\frac{1}{6}\gamma_{A}\gamma^{B}\omega^{Cn}_{A}\gamma_{Cn}\xi_{-}^{i}\,,
\eea
In the above, we have defined
\be
\cD^{3d}_{A}\xi_{+}^{i}=\nabla^{3d}_{A}\xi_{+}^{i}+V^{(3)}_{A}\sigma^{i}_{3\,j}\xi_{+}^{j},\quad\text{and}\quad V^{i}_{B\,j}=V^{(1)}_{B}\sigma^{i}_{1\,j}+V^{(2)}_{B}\sigma^{i}_{2\,j}\,.
\ee
Now, we require that the Killing spinor is such that 
\be
\xi_{+}^{i}\Big|_{r=\frac{\pi}{2}}=0
\ee
Using the fact that our squashed metric background is such that
\be
\omega^{Bn}_{A}=0,\qquad\text{for}\quad A,B=1,2,3\,, 
\ee
we get at $r=\frac{\pi}{2}$
\bea
\cD^{3d}_{A}\xi_{-}^{i}+T_{BC}(\gamma_{BC}\gamma_{A}+\frac{1}{3}\gamma_{A}\gamma_{BC})\xi_{-}^{i}=\frac{1}{3}\gamma_{A}\gamma^{B}\cD^{3d}_{B}\xi_{-}^{i}\,.
\eea
This is the 3-dimensional Killing spinor equation. Next, we consider the case of \eqref{NegativeProj.}. We get at $r=\frac{\pi}{2}$
\bea
&&\frac{1}{2}\omega^{Bn}_{A}\gamma_{Bn}\xi_{-}^{i}-2T_{Bn}(\gamma_{B}\gamma_{A}\gamma_{n}+\frac{1}{3}\gamma_{A}\gamma_{B}\gamma_{n})\xi_{-}^{i}=\frac{1}{6}\gamma_{A}\gamma^{B}\omega^{Cn}_{A}\gamma_{Cn}\xi_{-}^{i}
\eea
The above is the consistency condition the Killing spinor $\xi^{i}_{-}$ need to satisfy for the existence of supersymmetry on a manifold with boundary.
Using the fact that our squashed metric backgrounds are such that
\be
\omega^{Bn}_{A}=0,\qquad\text{for}\quad A,B=1,2,3\,, 
\ee
we obtain at $r=\frac{\pi}{2}$
\be
T_{Bn}(\gamma_{B}\gamma_{A}\gamma_{n}+\frac{1}{3}\gamma_{A}\gamma_{B}\gamma_{n})\xi_{-}^{i}=0
\ee
This should be true for every $A$. In particular, one possible solution of the above can be that the background is such that $T_{Bn}|_{r=\frac{\pi}{2}}=0$. This will not impose any condition of the Killing spinor.
\section{Background fields for the squashed geometry with $SU(2)\times U(1)$ isometry}\label{background}
\begin{equation}
\begin{split}\label{vt3t3b}
&v_{3,3}=-\frac{1}{h(r)}\left(1+\frac{h^{\prime}(r)}{2}\right)+\frac{h(r)}{2\sin^2{r}}+\frac{1}{2}\cot{\frac{r}{2}},\\
&t^{+}_{3}=\frac{1}{16 c\sin^2{\frac{r}{2}}}\left(2h^{\prime}(r)+\left(\frac{h(r)}{\sin{r}}\right)^2-\left(\cot{r}+\cot{\frac{r}{2}}\right)h(r)\right),\\
&t^{-}_{3}=\frac{c}{4h(r)\cos^2{\frac{r}{2}}}\biggl(h(r)-\sin{r}\biggr),\\
&M=2\left(\frac{h(r)}{\sin^2{r}}+\cot{\frac{r}{2}}\right)\frac{h^{\prime}(r)}{h(r)}+\left(\frac{h(r)}{\sin^2{r}}-2\cot{\frac{r}{2}}\right)
\frac{h(r)}{\sin^2{r}}+\left(\cot^2{r}-\cot^2{\dfrac{r}{2}}-2\right).
\end{split}
\end{equation}
\section{Supersymmetry transformations and vector multiplet Lagrangian}\label{SusyLagrangian}
First we list the SUSY transformations for the fields in the vector multiplet
\bea
&&\delta S=\bar\xi_{i}\lambda^{i}\,,\nn\\
&&\delta P=\bar\xi_{i}\gamma_{5}\lambda^{i}\,,\nn\\
&&\delta A_{\m}=\bar\xi_{i}\gamma_{\m}\lambda^{i}\,,\nn\\
&&\delta \vec D=i\vec \tau^{i}_{j}\bar\xi_{i}\gamma^{a}\mathcal D_{a}\lambda^{j}\,,\nn\\
&&\delta\lambda^{i}=\gamma^{a}\p_{a}S\xi^{i}-\gamma^{a}\p_{a}P\gamma_{5}\xi^{i}-\frac{1}{2}F_{ab}\gamma^{ab}\xi^{i}+2(ST_{ab}+\frac{1}{2}P\e_{abcd}T_{cd})\gamma^{ab}\xi^{i}-i\vec D\cdot\vec \tau^{i}{}_{j}\xi^{j}\nn\\
&&\qquad+\frac{1}{2}S\slashed \cD\xi^{i}-\frac{1}{2}P\slashed \cD\gamma_{5}\xi^{i}\nn\\
&&\delta\bar\lambda_{i}=\bar\xi_{i}\gamma^{a}\p_{a}S-\bar\xi_{i}\gamma_{5}\gamma^{a}\p_{a}P+\frac{1}{2}\bar\xi_{i}F_{ab}\gamma^{ab}-2\bar\xi_{i}(ST_{ab}+\frac{1}{2}P\e_{abcd}T_{cd})\gamma^{ab}+i\vec D\cdot\vec\tau^{j}{}_{i}\bar\xi_{j}\nn\\
&&+\frac{1}{2}S\mathcal\cD_{a}\bar\xi_{i}\gamma^{a}+\frac{1}{2}P\mathcal\cD_{a}\bar\xi_{i}\gamma^{a}\gamma_{5}
\eea
In the above $T_{ab}$ is a 2 form and not self or antiself dual i.e. $T_{ab}=T^{+}_{ab}+T^{-}_{ab}$ and
\bea
&&\mathcal D_{\m}\lambda^{i}=(\p_{\m}+\frac{1}{4}\omega_{\m ab}\gamma^{ab})\lambda^{i}+\frac{1}{2}\mathcal V^{i}_{\m\,j }\lambda^{j}\nn\\
&&\mathcal D_{\m}\xi^{i}=(\p_{\m}+\frac{1}{4}\omega_{\m ab}\gamma^{ab})\xi^{i}+\frac{1}{2}\mathcal V^{i}_{\m\,j }\xi^{j}
\eea
We will mention here the Lagrangian of the abelian theory. 
The action for the theory is
\be
I_V =  I_g + I_{\partial g}+ I_{\theta}
\ee
\bea
I_{g}&=&\frac{1}{2g^2}\int_{\cM} d^4x \Big( \frac{1}{2} F_{\mu\nu}F^{\mu\nu}-8F^{\m\n}(ST_{\m\n}+\frac{1}{2}P\epsilon_{\m\n\rho\eta}T^{\rho\eta})+ \overline{\lambda_i}\slashed{\cD} \lambda^i \nn\\
&&+ (\partial_\mu P)^2- (\partial_\mu S)^2 - \vec{D}^2 -\frac{M}{2}(P^2 - S^2)+16(S^{2}+P^{2})T_{ab}T^{ab}+16SP \epsilon_{abcd}T^{ab}T^{cd}\Big),\nn\\
\eea
\begin{eqnarray}
I_{\partial g}&=& \frac{1}{g^{2}}\int_{\partial \cM} d^3 x \left( \frac{i}{4}\tau_3{}^i{}_{j}\, \overline{\lambda_{i}}\gamma_5 \lambda^j + P (D_3 -\partial_n P)\right)\,,
\\
I_\theta &=& \frac{ \theta}{4\pi^2} \left[ \frac{i}{4} \int_{\cM} d^4 x F_{\mu\nu}\widetilde{F}^{\mu\nu} +\int_{\partial \cM}d^3 x\left( \frac{1}{2} \tau_3{}^i{}_{j}\,\overline{\lambda_-}_i \lambda^j_-  - i S(D_3 -\partial_n P )   \right)\right]\,.\nn\\
\end{eqnarray}
\section {Functions}\label{Functions}
The explicit form of the function that appears in the expression for $\tau_{R}$~\eqref{NonChiralLargeCoupling} is
\bea
f_{1}(x)&=&\frac{e^{\frac{i}{2}(3\alpha+\frac{\pi}{2})}}{18\pi(1+2\cosh 2x)^{2}}\Big[4\pi^{2}(\cosh 4x-3\cosh 2x-7)-3\sqrt{3}\pi((1+2\cosh2x)(-3+2a\sinh \alpha)\nn\\
&&+20x\sinh 2x)+18x(3x-2a\sinh \alpha(\sinh 2x+4\sinh 4x))\Big]\,,
\eea
where $a=\frac{54\sqrt{3}-90\pi+8\sqrt{3}\pi^{2}}{9\Big(8\pi(3\sqrt{3}-2\pi)-27\Big)}$.

The explicit form of the function that appears in the expression for $\tau_{R}$~\eqref{TauRChiral} is
\bea
g(x,\alpha')&=&\frac{e^{i\frac{\pi}{24}}}{64\sqrt{2}\pi \tau'}e^{2x}(1+e^{2x})^{-\frac{5}{2}+\frac{ix}{\pi}}\frac{1}{\sinh^{3}x}\Big(-8(\pi+i x)\frac{x}{\sinh x}\nn\\
&&-8\cosh x (\pi+\pi x-2i x^{2}+2 e^{i\alpha'}(i\pi+2x-2x\coth x)\sin\alpha'\nn\\
&&+2(3i\pi^{2}+4\pi x-8ix^{2})\sinh x\Big)(3\cosh2x-3-8x+4\sinh 2x)\,.
\eea
%%%%%%%%%%%%%%%%%%%%%%%%%%%%%%%%%%%%%%%%%
%\bibliography{supergraphene}
\providecommand{\href}[2]{#2}\begingroup\raggedright\endgroup

\end{document}